\documentclass{aa}

\usepackage{gensymb}

\usepackage{changepage}

\usepackage{graphicx}
\usepackage{txfonts}
\usepackage[colorlinks = true,
            linkcolor = blue,
            urlcolor  = blue,
            citecolor = blue,
            anchorcolor = blue]{hyperref}

\begin{document}

   \title{Occurrence rate of hot Jupiters orbiting red giant stars}

   \author{Milou Temmink
          \inst{1}
          \and
          Ignas A.G. Snellen\inst{1}
          }

   \institute{1: Leiden Observatory, Leiden University, Postbus 9513, 2300 RA, Leiden, The Netherlands, \\
   \email{temmink@strw.leidenuniv.nl}
             }

    \date{}
   \date{Received ..., 202x; accepted ..., 202x}
 
  \abstract
   {Hot Jupiters form an enigmatic class of object with yet unclear formation pathways. Determination of their occurrence rates as function orbit, planet and stellar mass, and system age,  can be an important ingredient for understanding how they form. To date, various Hot Jupiters have been discovered orbiting red giant stars and deriving their incidence would be highly interesting.}
   {In this study we aim to determine the number of Hot Jupiters in a well-defined sample of red giants, estimate their occurrence rate and compare it with that for A, F and G-type stars. }
   {A sample of 14474 red giant stars, with estimated radii between 2 and 5 $R_\odot$, was selected using Gaia to coincide with observations by the NASA TESS mission. Subsequently, the TESS light curves were searched for transits from Hot Jupiters. The detection efficiency was determined using injected signals, and the results further corrected for the geometric transit probability to estimate the occurrence rate.}
   {Three previously confirmed Hot Jupiters were found in the TESS data, in addition to one other TESS Object of Interest, and two M-dwarf companions. This results in an occurrence rate of $0.37^{+0.29}_{-0.09}$\%. Due to the yet large uncertainties, this cannot be distinguished from that of A-, F- and G-type stars.. We argue that it is unlikely that planet engulfment in expanding red giants plays yet an important role in this sample.}
   {}


   \maketitle
%

\section{Introduction}
\noindent Hot Jupiters (HJs) are very interesting objects to study. The close proximity to their host star makes them perfect candidates to observe with the transit method and causes their atmospheres to reach up to similar temperatures as some M-dwarf stars. However, it is still unknown how these objects form. To date, various hypotheses have been develop which could possibly explain their formation. The most promising are in situ formation, disk migration and high eccentricity tidal migration \citep{DJ18}. \\
\indent One way to possibly gain more information on the formation and the evolution of HJs is through the study of the occurrence rates as function of planet and stellar mass, orbits and system-ages. \cite{Zetal19} determined the HJ occurrence rate to be $0.43\pm0.15$ and $0.71\pm0.31\%$ for F- and G-type stars, respectively, while that around Kepler stars was estimated to be $0.43\pm0.05$ \citep{Fetal13}, $0.43^{+0.07}_{-0.06}$ \citep{Metal17} and $0.57^{+0.14}_{-0.12}$ \citep{Petal18}. \cite{Zetal19} found the occurrence rate around A-stars to be $0.26\pm0.11$. Furthermore, \citet{GrunblattEA19} have estimated the occurrence rate of giant planets around low-luminosity red giants to be $0.51\pm0.29$\%.\\
\indent It is particularly interesting to determine the occurrence rate of HJs around red giant stars. As evolved stars have gone through their initial hydrogen core burning phase, an estimate for the occurrence rate of HJs may tell us more about the evolution of these planetary systems and the survivability of close-in planets. To date, various HJs have been found to orbit red giants, Kepler-91b \citep{LBetal14} being one of the most prominent. \\
\indent Besides hot Jupiters other kinds of exoplanetary populations have been found orbiting red giant stars as well. For example, both hot Neptunes (e.g. K2-39b, \citet{vEylenEA16}) and warm eccentric Jupiter (e.g. Kepler-432b, \citet{CiceriEA15, QuinnEA15}) subpopulations have been found besides the hot Jupiter one. In addition, the detections of Kepler-56 b and c have shown that red giants can contain multiplanetary systems \citep{BoruckiEA11, HuberEA13}. \\
\indent In this paper we aim to estimate the occurrence rate of HJs around red giants. In Section \ref{sec:S2} we describe how we created a well-defined sample of red giants in the solar neighbourhood observed by the NASA TESS mission, while in Section \ref{sec:S3} we explain how we searched for HJs transiting these stars. Section \ref{sec:S4} describes how we estimated the occurrence rate of HJs around red giants by combining the geometric probability of observing transits and the fraction of detected HJs. Our discussions and conclusions can be found in Section \ref{sec:S5}.

\section{Sample selection and \textit{TESS} observations} \label{sec:S2}
\subsection{GAIA sample of red giant stars}
\begin{figure}[ht!]
    \includegraphics[width=\hsize]{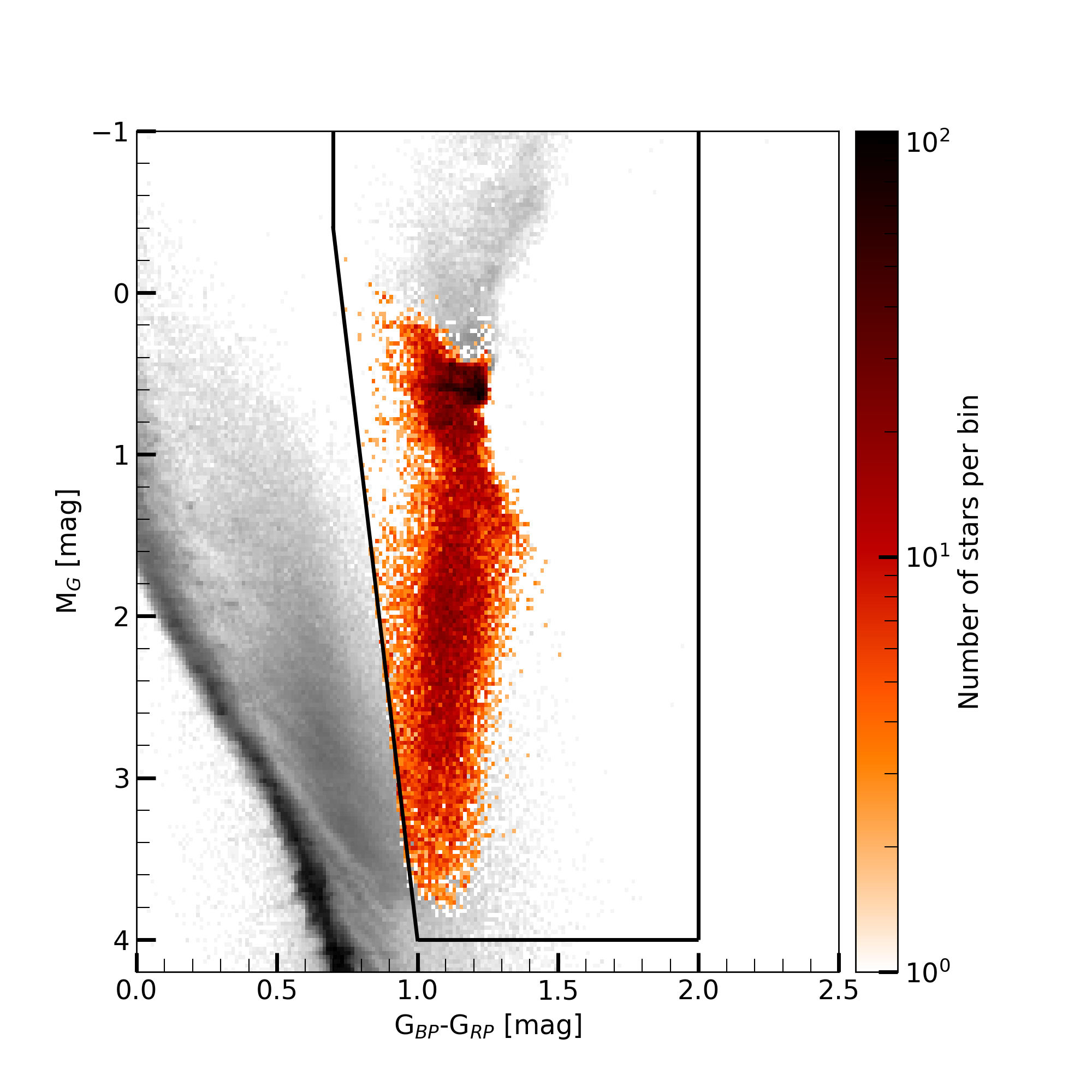}
    \caption{The colour-magnitude diagram of stars which have survived the imposed selection criteria with the black box representing the boundaries of the region enclosing the RGB. The colour ($G_\textnormal{BP}-G_\textnormal{RP}$) of the stars can be found on the x-axis, while the y-axis denotes the absolute magnitudes. The logarithmic colour-map indicates the number of stars in each two-dimensional bin. stars removed by the selection criteria are shown in grey.}
    \label{fig:HRD}
\end{figure}

All stars with available photometric measurements in the $G$, $G_\textnormal{BP}$ and $G_\textnormal{RP}$, passbands, with estimated absolute magnitudes of $M_\textnormal{G}\leq5.5$ and with parallaxes of $\varpi\geq2$ milliarcseconds was selected using the \textit{GAIA} DR2 catalogue \citep{GC16}. To ensure that the sample contained only stars with sufficiently precise photometric and astrometric properties, the sample was subject to the following selection criteria, similar to those imposed in \cite{GC18HRD}:
\begin{itemize}
    \item All stars with high astrometric excess noise were removed by imposing the following constraint,
    \begin{align}
        \sqrt{\chi^2/(\nu'-5)} < 1.2\max(1,\exp\left(-0.2(G-19.5))\right),
    \end{align}
    where $\chi^2$ is the \textsc{astrometric\_chi2\_al} parameter and $\nu'$ is the \textsc{astrometric\_n\_good\_obs\_al} parameter. The values for $\chi^2$ and $\nu'$ were obtained from the \textit{GAIA} DR2 catalogue.
    \item Only stars with a precision of at least 10\% on their measured parallax were retained in the sample,
    \begin{align}
        \frac{\varpi}{\sigma_\varpi}\geq 10.
    \end{align}
    to be able to correctly estimate the absolute magnitudes.
    \item We ensured that the stars have accurate photometry quantities by imposing the following constraints on the signal-to-noise in the different \textit{GAIA} passbands,
    \begin{align}
        \frac{F_\textnormal{G}}{\sigma_{F_\textnormal{G}}} \geq50, \rule{0.2cm}{0pt}\frac{F_\textnormal{BP}}{\sigma_{F_\textnormal{BP}}}\geq20, \rule{0.2cm}{0pt}\frac{F_\textnormal{RP}}{\sigma_{F_\textnormal{RP}}}\geq20.
    \end{align}
    \item The \textit{GAIA} $G_\textnormal{BP}$ and $G_\textnormal{RP}$ have not been corrected for blending of background sources. To remove the stars whose fluxes might suffer from contamination from background sources, all stars with \textsc{phot\_bp\_rp\_excess\_factor} $\left(C=\frac{F_\textnormal{BP}+F_\textnormal{RP}}{F_\textnormal{G}}\right)$ falling outside the following two limits were removed from the collection,
     \begin{align}
        C & > 1.0 + 0.015(G_\textnormal{BP}-G_\textnormal{RP})^2, \nonumber\\
        C & < 1.3 + 0.06(G_\textnormal{BP}-G_\textnormal{RP})^2.
    \end{align}
    \item Additionally, we ensured that all stars have well measured astrometric quantities by removing those with the Re-normalised Unit Weight Error above 1.4.
\end{itemize}
\indent We reduced the sample further by ensuring that all stars have been observed during the first two years of \textit{TESS} observations and for which it is expected that transits can be detected:
\begin{itemize}
    \item We utilised the python-package \textsc{TESS-point} to check if the stars, based on their positional coordinates (right ascension and declination), are located in one of the first 26 \textit{TESS} sectors. 
    \item As the transit depth scales inversely with the stellar radius squared, an upper limit of $R\leq10R_\odot$ was imposed on the stellar radii. Furthermore, a lower limit of $R\geq2R_\odot$ was imposed, as we assumed this to be the typical stellar radius at the base of the red giant branch (RGB).
    \item For faint stars, transit signals might get lost in the noise. We kept the stars with apparent magnitudes of $T\leq10$ in our sample. The apparent magnitudes were estimated using equation 1 of \cite{Setal19},
    \begin{align}
        T = & G - 0.00522555(G_\textnormal{BP}-G_\textnormal{RP})^3 + \nonumber \\ 
        & 0.0891337(G_\textnormal{BP}-G_\textnormal{RP})^2 - 0.633923) + 0.032447.
    \end{align}
\end{itemize}
\indent Finally, to remove stars from the sample that do not reside on the red giant branch, stars were selected based on their position in the colour- magnitude diagram (CMD). This criteria ensures that massive subgiant stars of similar radii do not contaminate the sample of red giant stars. The selection region has the following boundaries,
\begin{align}
    G_\textnormal{BP}-G_\textnormal{RP} & \geq (M_\textnormal{G} + 11.13)/15.33, \nonumber \\
    G_\textnormal{BP}-G_\textnormal{RP} & \leq 2.0, \\
    1.0 \leq M_\textnormal{G} & \leq 4.0. \nonumber 
\end{align}
The resulting CMD of the stars surviving the imposed selection criteria, together with the selection boundaries, is displayed in Figure \ref{fig:HRD}. \\
\indent The final sample consisted of 35250 red giants.

\subsection{\textit{TESS} observations}
If available, the 2-minute cadence \textit{TESS} 'Pre-search Data Condition SAP' (PDCSAP; \cite{Jetal16}) lightcurves were downloaded, otherwise the lightcurves were extracted from the 30-minute full frame images (FFIs) using the python-package \textsc{tesseract}\footnote{\textit{tesseract}: \url{https://github.com/astrofelipe/tesseract}}. About 83\% of the lightcurves were obtained using \textsc{tesseract} and all these lightcurves have obtained using the default aperture option, which is based on the K2P2 method (see \citet{K2P2}). \\
\indent The baselines of the lightcurves were detrended using an iterative spline fitting method \citep{VJ14}. The splines consisted of polynomials of degree 3 and in each iteration we excluded the $3\sigma$ outliers, until no outliers remained. We further cleaned the lightcurves by removing all data points lying $3\sigma$ above the detrended baselines. Following \cite{Vetal16} the two lowest data points were removed as well.

\section{BLS search for hot Jupiters} \label{sec:S3}
To search the cleaned lightcurves for transits arising from hot Jupiter candidates on periods between 4 and 25 days, we used the  BoxLeastSquares (BLS) implementation of the python-package \textsc{Astropy} \citep{Astropy1, Astropy2}. The periodograms, yielded by the BLS method, were normalised following the steps given in \cite{Vetal16}. First, we subtracted the changing baseline level of the periodograms, to avoid spurious detections at longer periods. Next, we converted the BLS power measurements to signal-to-noise ratios by using equation 4 of \cite{Vetal16}, 
\begin{align}
    \textnormal{SNR} = \frac{P_\textnormal{BLS}}{\textnormal{MAD}/0.67}.
\end{align}
Here, MAD stands for the median absolute deviation. \\
\indent To validate any signals arising in the normalised periodograms, the periodograms were subject to a self-imposed algorithm, consisting of the following steps:
\begin{enumerate}
    \item We set a limit of $\textnormal{SNR}\geq9$ on the highest peak found in the periodogram. If the highest peak did not meet this limit, it was assumed that the lightcurve did not contain any transits. Subsequently, the algorithm would consider the combination of transit parameters (orbital period, midpoint of first transit, transit depth and transit duration) belonging to the signal yielding the highest peak in periodogram were determined.
    \item If the corresponding transit duration was found to be longer than 25\% of the orbital period, the found signal was rejected. Also, in the case only a single data point was found in the fitted boxes of the BLS method, the signal was considered to be a false positive. In these cases, the algorithm would remove the data points yielding the signal from the lightcurves, generate a new periodogram and return to step 1. 
    \item The final step estimated the radius of the candidate, using
    \begin{align}
        R_\textnormal{object} = \sqrt{\Delta F}R,
    \end{align}
    here $\Delta F$ is the transit depth and $R$ is the stellar radius. The candidate was assumed to be a potential hot Jupiter if $R\leq2.2R_\textnormal{Jup}$.
\end{enumerate}
The periodograms which passed all the steps were subsequently subjected to an individual visual inspection. The visual inspection allowed to distinguish between real candidates and false positives. Among false positives, we included binary stars, features in the lightcurves which clearly did not have a transit shape and objects that were found to transit background stars. The objects orbiting background stars were investigated by thoroughly studying the \textit{TESS} target pixel files.

\subsection{Transit search results}
The search for hot Jupiter candidates lead to 1392 potential detections, of which all except 6 were rejected upon visual inspection. The other 1386 detections were found to be false positives, binary stars or objects orbiting stars in the background of the target star. The latter were found by thoroughly studying the \textit{TESS} target-pixel files and the lightcurves associated with each pixel. \\
\indent The potential candidates were found orbiting the stars CD-25 2180, HD 1397, HD 93396 (also known as KELT-11), HD 221416, TYC 4004-1837-1 and TYC 5456-76-1. A literature study revealed that the candidates orbiting HD 1397, KELT-11 and HD 221416 correspond to the already confirmed exoplanets HD 1397b \citep{Netal19}, KELT-11b \citep{Petal17} and HD 221416b \citep{Hetal19}. The candidate orbiting the star TYC-5456-76-1 was found to correspond to the TESS object of interest, TOI-2669.01, and the planetary nature of this candidate has recently been confirmed by \cite{GrunblattEA22}. The final two candidates orbiting CD-25 2180 and TYC-4004-1837-1 were found to be M-dwarf stellar companions, using existing radial velocity measurements. \\
\indent As mentioned before, 83\% of the lightcurves have been obtained using \textsc{tesseract}. The lightcurves of TYC-5456-76-1, CD-25 2180 and both lightcurves of TYC-4004-1387-1 have obtained using \textsc{tesseract}. The lightcurves of KELT-11 and HD 221416 were both obtained from the SPOC pipeline, while one of the lightcurves of HD 1397 has been obtained using the SPOC pipeline (sector 1) and the other one has been obtained using \textsc{tesseract} (sector 2). As both pipelines have resulted in a similar amounts of observed candidates, both pipeline appear to reach similar sensitivities.

\subsection{Fitting the transits}
To extract as much information about the found objects and their orbits, we fitted a transit model to the lightcurves using a Markov Chain Monte Carlo (MCMC) algorithm, using the python-package \textsc{emcee} \citep{FMetal14}. The lightcurves were first re-detrended, using a Gaussian Processes method. The Gaussian Processes (GP) method was implemented using the \textsc{Juliet} \citep{Eetal19} package. To ensure that the transit would not influence the GP trend, the out-of-transit data was removed by phase-folding the lightcurves, based on the period and midpoint-of-first-transit found with our BLS search, and excluding all points which fall within 0.02 of the absolute phases. Subsequently, the GP trend was determined using the default Matern kernel, for which the used hyperparameters can be found in Table \ref{tab:GP_HP}. The detrended lightcurves can be found in Appendix \ref{App:A}.\\
\indent The model lightcurves were created using the python-package \textsc{PyTransit} \citep{Petal15}, assuming for simplicity linear limb-darkening with a value of 0.5 and circular orbits. The remaining parameters (orbital period, midpoint of first transit, ratio of object radius over stellar radius, semi-major axis in stellar radii and the inclination) were fitted with the MCMC algorithm. The a priori distributions for the various parameters are in Table \ref{tab:UD_TF}. The best fitting transit models and the corresponding values for the transit parameters can be found in appendix \ref{App:B}. \\
\begin{table}[ht!]
    \centering
    \begin{tabular}{cc}
        \hline
        \hline
        Parameter & Uniform distribution \\
        \hline
        $P$ [days] & U($P_\textnormal{BLS}$-0.05,$P_\textnormal{BLS}$+0.05) \\
        $T_0$ [TJD] & U($T_{0,\textnormal{BLS}}($-0.05,$T_{0,\textnormal{BLS}}$+0.05) \\
        $k$ & U($k_\textnormal{BLS}/2, 2k_\textnormal{BLS}$) \\
        $a/R$ & U(1,20) \\
        $i$ [degrees] & U(60, 90) \\
        \hline
    \end{tabular}
    \caption{The a-priori parameter limits used in the MCMC algorithm. A subscript 'BLS' indicates that the value obtained from the BLS method has been used.}
    \label{tab:UD_TF}
\end{table}

\begin{table}[ht!]
    \centering
    \begin{tabular}{ccc}
        \hline
        \hline
        Parameter & Distribution & Hyperparameters \\
        \hline
        $m_\textnormal{dilution}$ & Fixed & 1.0 \\
        $m_\textnormal{flux}$ & Normal & (0.0, 0.1) \\
        $\sigma_\omega$ & Loguniform & (1e-6, 1e6) \\
        $\sigma_\textnormal{GP}$ & Loguniform & (1e-6, 1e6) \\
        $\rho$ & Loguniform & (1e-3, 1e3) \\
        \hline
    \end{tabular}
    \caption{The distributions and hyperparameters used during the Gaussian Processes method. Here, $m_\textnormal{dilution}$ is a dilution factor and $m_\textnormal{flux}$ is the mean out-of-transit flux. $\sigma_\omega$ is a (unknown) jitter term, $\sigma_\textnormal{GP}$ is the amplitude of the GP and $\rho$ denotes the Matern time-scale.}
    \label{tab:GP_HP}
\end{table}

\section{Occurrence rate} \label{sec:S4}
To estimate the occurrence rate of HJs around red giants, we derived the geometric probability and the probability of actually detecting transits of the planets as function of orbital period and planetary and stellar sizes.

\subsection{Geometric probability}
The geometric probability is the probability of observing a transit in a randomly oriented system. It depends mostly on the ratio of the semi-major axis of the orbit ($a$) over the stellar radius. The geometric probability can be calculated using equation 2 of \cite{DK2014}:
\begin{align}
    P_\textnormal{geometric} = \left(\frac{R}{a}\right)\left(\frac{1+e\sin(\omega)}{1-e^2}\right) \stackrel{e=0}{=} \frac{R}{a},
\end{align}
where after the last equality sign the orbits, for simplicity, have been assumed to be circular, by setting the eccentricity to $e=0$, and are shown in Table \ref{table:GeoProb}. \\
\indent For our occurrence rate estimate, we used the average geometric probability of $0.196^{+0.014}_{-0.057}$.

\begin{table*}
\caption{The stellar masses, stellar radii, planetary radii, planetary masses and the semi-major axes used in the calculation for the geometric probability. The calculated geometric probability of detecting a hot Jupiter orbiting the star at the given semi-major axis is shown as well.}    
\label{table:GeoProb}     
    \begin{adjustwidth}{-0.7cm}{}   
    \begin{tabular}{c c c c c c c c }     
    \hline\hline       
    Star & M [$M_\odot$] & R [$R_\odot$] & $R_p$ [$R_\textnormal{Jup}$] & $M_p$ [$M_\textnormal{Jup}$] & a [AU] & Geom. probability & References \\ 
    \hline                    
       HD 1397 & $1.322^{+0.049}_{-0.042}$ & $2.328^{+0.050}_{-0.056}$ & $1.026\pm0.026$ & $0.415\pm0.020$ & $0.1097^{+0.011}_{-0.013}$ &  $0.099\pm0.003$ & 1,2 \\
       KELT-11 & $1.438^{+0.061}_{-0.052}$ & $2.72^{+0.21}_{-0.17}$ & $1.30^{+0.15}_{-0.12}$ & $0.195\pm0.018$ & $0.06229^{+0.00088}_{-0.00076}$ &  $0.203^{+0.18}_{-0.15}$ & 3 \\
       HD 221416 & $1.212\pm0.074$ & $2.943\pm0.064$ & $0.836^{+0.031}_{-0.028}$ & $0.190\pm0.018$ & $0.1228^{+0.0025}_{-0.0026}$ &  $0.111\pm0.005$ & 4 \\
       TYC 5456-76-1 & $1.19\pm0.16$ & $4.10\pm0.04$ & $1.76\pm0.16$  & $0.61\pm0.19$ & \tablefootmark{a}$0.05^{+0.07}_{-0.01}$ & $0.372^{+0.053}_{-0.226}$ & 5 \\
    \hline                  
    \end{tabular}
    \tablebib{(1)~\citet{Netal19}; (2)~\citet{Betal19}; (3)~\citet{Petal17}; (4)~\citet{Hetal19}; (5)~\citet{GrunblattEA22}}
    \end{adjustwidth}
\end{table*}

\subsection{Observational probability}
\begin{figure*}[h]
    \resizebox{\hsize}{!}{\includegraphics{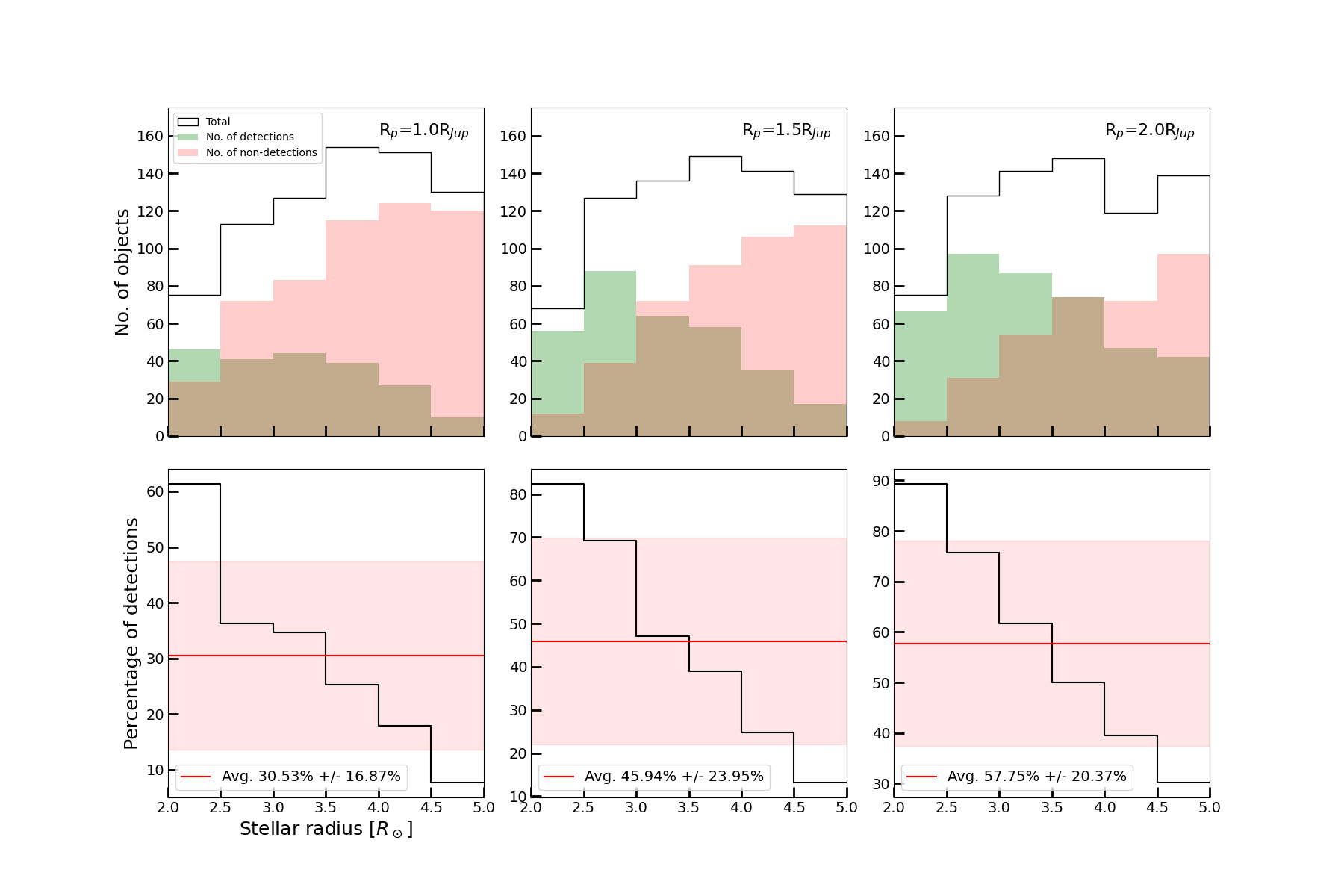}}
    \caption{The results from the injecting artificial transit test. Top row: The left panel shows the result for injected planets with radius of $R_p=1.0R_\textnormal{Jup}$, the middle panel for planets of radius $R_p=1.5R_\textnormal{Jup}$ and the right panel for planets with radius of $R_p=2.0R_\textnormal{Jup}$. The green shaded areas shows the correctly detected artificial transits for different stellar radii, while the red shaded areas show the non-detections. The black line shows the total amount of injected planets for the different stellar radii. Bottom row: The percentages of correctly detected artificial transits for stellar radii in the range $2.R_\odot\leq R\leq5R_\odot$. The red line indicates the average percentage and the red shaded area indicates the $1\sigma$-errors.}
    \label{fig:ObsPer}
\end{figure*}

To estimate the fraction of hot Jupiters detected, we performed artificial injection tests. The injected transits were based on the orbit of Kepler-91b \citep{LBetal14}. We carried out the injection tests for planetary radii of $R_p=1.0R_\textnormal{Jup}$, $1.5R_\textnormal{Jup}$ and $2.0R_\textnormal{Jup}$, assuming an inclination of 90\degree. Our injected planets have periods between 4 and 15 days, randomly picked from a log-normal distribution. This period range covers the same range as for the detected planets in the geometric probability. For each planetary radius we conducted 5 different tests, where in each test 150 stars were randomly chosen from our sample of red giants. \\
\indent Figure \ref{fig:ObsPer} shows the number of correctly detected and non-detected injected transits (top row), with the percentages of the correctly detected transits (bottom row). \\
\indent As the confirmed exoplanets all orbit red giants with stellar radii of $R<5R_\odot$, we only focus on these stellar radii in our calculation for the observational probability, to ensure that both probabilities (geometric and observational) account for a similar population of red giant stars. This means that in the subsequent calculation of the total occurrence rate, we will only consider the 14474 stars in our sample which have a radius of  $R<5R_\odot$. To account for the fact that more smaller hot Jupiters with radii $1.0R_\textnormal{Jup}\leq R \leq 1.5R_\textnormal{Jup}$ have been observed than the larger hot Jupiters ($R>1.5R_\textnormal{Jup}$), we have added weights based on the observed hot Jupiters with radii $1.0R_\textnormal{Jup}\leq R\leq2.0R_\textnormal{Jup}$ and orbital periods $P\leq15$ days\footnote{The observed hot Jupiters have been acquired from Exoplanet.eu (\url{http://exoplanet.eu/}, on July 18, 2022) using the query: \textsc{radius:rjup $>=$ 1 AND radius:rjup $<=$ 2 AND period:day $<=$ 15}}. \\
We acquired a total number of 501 observed hot Jupiters, of which 323 have radii between $1.0R_\textnormal{Jup}\leq R \leq 1.33R_\textnormal{Jup}$, 136 have radii between $1.33R_\textnormal{Jup}\leq R \leq 1.66R_\textnormal{Jup}$ and 42 have radii between $1.66R_\textnormal{Jup}\leq R \leq 2.0R_\textnormal{Jup}$. Based on these values, we establish that about 2 times more hot Jupiters have radii close to $1.0R_\textnormal{Jup}$ compared to $1.5R_\textnormal{Jup}$ and 3 times more hot Jupiters have radii close to $1.5R_\textnormal{Jup}$ compared to $\sim2.0R_\textnormal{Jup}$. Subsequently, we determined our fraction of observed hot Jupiters using a weighted average, were we have used weights of $2$, $1$ and $\frac{1}{3}$, respectively. \\
\indent Our weighted averaged yields a percentage of $0.38\pm0.13$ detected hot Jupiters.

\subsection{Total probability and occurrence rate}
Combining the geometric and detection probability, we find the combined probability to see a hot Jupiter with our method to be $0.07\pm0.03$. In our search for HJs orbiting red giants, we find 4 systems. Correcting for the geometric probability and detection fraction, we estimate there to be 40-96 planets orbiting our sample stars. This yields for our sample of 14474 red giants with $R\leq5R_\odot$ an occurrence rate of $0.37^{+0.29}_{-0.09}$.

\section{Discussion} \label{sec:S5}
\subsection{Occurrence rate}
The occurrence rate derived here of $0.37^{+0.29}_{-0.09}$\% agrees, within the uncertainties, with those for the A-, F- and G-type stars of, respectively, $0.26\pm0.11$, $0.43\pm0.15$ and $0.71\pm0.31\%$ \citep{Zetal19}. These occurrence rates were derived for hot Jupiters with orbital periods between 0.9 and 10 days. No clear distinction can yet be made between the occurrence rates of these different stellar types. The occurrence rate also agrees with that for hot Jupiters orbiting the \textit{Kepler}-stars of $0.43\pm0.05$\% ($0.8\leq P\leq10$ days, \citet{Fetal13}), $0.43^{+0.07}_{-0.06}$ ($P\leq10$ days, \citet{Metal17}) and $0.57^{+0.14}_{-0.12}$ ($1\leq P\leq10$ days, \citet{Petal18}). Finally, our found value is similar to the one reported by \citet{GrunblattEA19} for giant planets orbiting low-luminosity red giants ($P\leq10$ days) in the \textit{K2}-observations, $0.51\pm0.29$\%. All-in-all the found occurrence rate for hot Jupiters orbiting red giant stars cannot (yet) be distinguished from the occurrence rates reported for the aforementioned stellar types. \\
\indent The masses of the three host stars already discussed in the literate are estimated to be in the range $1.2-1.5 M_\odot$, meaning that they used to be F-type main-sequence stars. As mentioned above, \cite{Zetal19} found an occurrence rate of $0.43\pm0.15$ for these types of stars. A larger sample and/or more \textit{TESS} data are needed to investigate potential differences between these populations. \\
\indent In addition, once more hot Jupiters have been found orbiting red giants, the occurrence rate can also be  improved by including the nonzero eccentricities of the detected planets in the calculation of the geometric probability and in the injection-retrieval tests.

\subsection{Possible planet engulfment}
\begin{figure*}[h]
   \resizebox{\hsize}{!}{\includegraphics{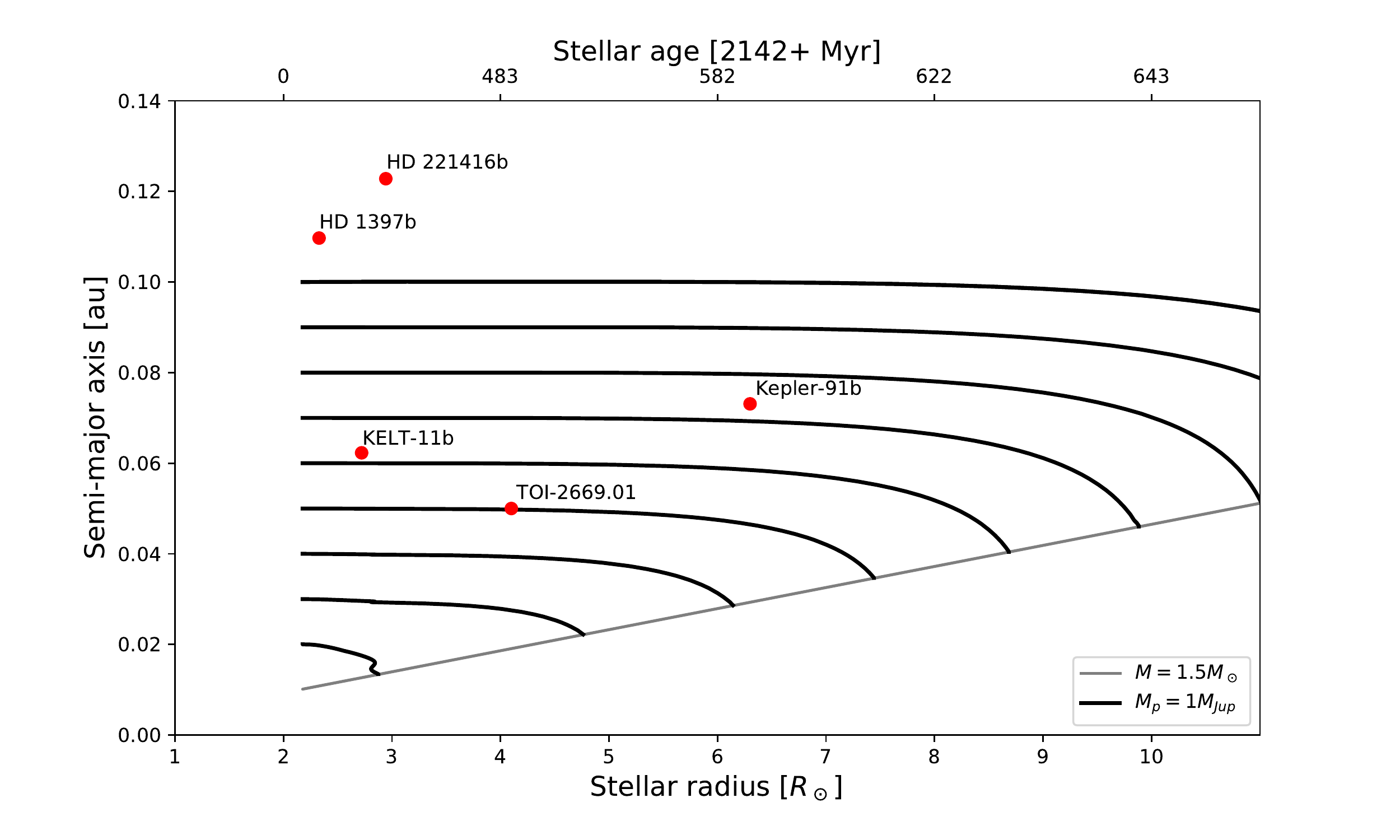}}
    \caption{The expected orbital evolution of Jupiter-mass planets orbiting a red giant of mass $1.5M_\odot$, on orbits with initial semi-major axis within the range of 0.02 AU to 0.10 AU, following \cite{Ketal11}. The HJs in our sample are indicated with Kepler-91 included for comparison.}
    \label{fig:PlEn}
\end{figure*}

A possible reason that might explain a lower occurrence rate of HJs orbiting red giants and the low number of planets that have been detected, is planet engulfment \citep{LBetal16}. Planet engulfment follows from the tidal dissipation of the orbit of the planet, due to interactions between the planet and the convective envelope of the red giant. \\
\indent We, however, argue that planet engulfment is not yet of importance during these early stages of the red giant phase. Figure \ref{fig:PlEn} shows the orbital evolution of Jupiter-mass planets, on various initial semi-major axes, during the early red giant phase of a star of mass $1.5M_\odot$. \\

\indent The evolution was modelled using the Modules for Experiments in Stellar Astrophysics code (\textit{MESA} v15140; \cite{Paxton2011}, \cite{Paxton2013}, \cite{Paxton2015}, \cite{Paxton2018} and \cite{Paxton2019}). The star was evolved from the pre-main sequence phase up to the core helium burning phase, the end of the RGB phase. During the evolution, the mass loss was estimated using Reimers' empirical formula \citep{Reim1975}: 
\begin{align} \label{eq:RMF}
    \dot{M} = -4\times10^{-13}\eta\frac{L}{L_\odot}\frac{R}{R_\odot}\frac{M_\odot}{M}\textnormal{ $M_\odot$/yr}.
\end{align}
Here, $L$, $M$ and $R$ are the stellar luminosity, mass and radius, respectively, all given in solar units. $\eta$ is a parameter of order unity, whose value usually varies between $0.2$ and $1.0$. A lower value of $\eta$ is often used for metal-poor stars, as the coupling of photons to the stellar gas is lower the fewer the heavy elements are present \citep{KWW12}. In the simulations we have used a fixed value of $\eta=0.5$. \\
\indent The orbital decay of the planet was modelled using the rate of change of the semi-major axis (equation 1 of \cite{Ketal11}):
\begin{align} \label{eq:RoC}
    \left(\frac{\dot{a}}{a}\right) = -6\frac{k}{\tau_d}\frac{M_p}{M}\left(1+\frac{M_p}{M}\right)\left(\frac{R}{a}\right)^8-\frac{\dot{M}}{M}.
\end{align}
Here, $M_p$ denotes the mass of the orbiting planet and $a$ is the semi-major axis of the orbit. $k$ is a dimensionless factor, which can be calculated using equation 2 of \cite{Ketal11},
\begin{align}
    k = \frac{f}{6}\frac{M_\textnormal{env}}{M}\rule{1cm}{0pt}\textnormal{where}\rule{1cm}{0pt}f=\min\left[1,\left(\frac{P}{2\tau_d}\right)^2\right].
\end{align}
$M_\textnormal{env}$ is the mass that is contained within the convective envelope of the RGB star, while $P$ denotes the orbital period. The orbital period was calculated using Kepler's third law,
\begin{align} \label{eq:K3L}
    P^2 = \frac{4\pi^2}{GM}a^3.
\end{align}
Furthermore, $\tau_d$ is the eddy turnover timescale, which was calculated using equation 4 of \cite{Ketal11},
\begin{align}
    \tau_d = \left[\frac{M_\textnormal{env}(R-R_\textnormal{env})^2}{3L}\right]^{\frac{1}{3}},
\end{align}
where $R_\textnormal{env}$ is the radius at the base of the convective envelope. The simulation of the orbital decay was halted as soon as the semi-major axis became smaller than the stellar radius. \\
\indent As can be seen, the orbital decay for different initial semi-major axes becomes of importance once the star is starting filling up a large portion of the planetary orbit. None of the confirmed exoplanets in this sample are experiencing significant orbital decay and they are, currently, not in danger of getting engulfed. For Kepler-91b, on the other hand, orbital decay might become important within tens of Myr \citep{LBetal14}, which is also visible in Figure \ref{fig:PlEn}. In addition, ultra-hot Jupiters (periods $<2$ days), which account for a small sub-sample of the HJ population, will likely not have survived this early red-giant phase. \\ 
\indent It must be noted that orbital decay becomes is more important for heavier planets around less massive stars, as can be inferred from Equation \ref{eq:RoC}. In addition, the confirmed exoplanets detected in this work shown in Figure \ref{fig:PlEn} are all less massive than $1 M_\textnormal{Jup}$ (see Table \ref{table:GeoProb}). Subsequently, their orbital decay will be even less severe than what is suggested in Figure \ref{fig:PlEn}. 

\subsection{Future searches}
The occurrence rate estimate as presented in this work is only based on a few objects and may be prone to biases due to the detection probability being a strong function of stellar apparent magnitude and stellar radius. Since \textit{TESS} continues to observe, the available data to search for HJs around red giant stars will only increase and mitigate this issue. It may also help to conduct a more sensitive study using Difference Imaging Analysis \citep{OS18,Metal20} or a Principle Component Analysis (see, for example, the eleanor \citep{eleanor} or the Giants \citep{NPLB} pipelines) to the light curves obtained from the FFIs. These pipelines could yield cleaner light curves, an increase in detection probability and a more precise estimate for the occurrence rate. For example, our method has missed one possible candidate, TOI-4551.01\footnote{\textit{ExoFOP}-page of TOI-4551.01: \url{https://exofop.ipac.caltech.edu/tess/target.php?id=204650483}}, due to transit-like signals in the lightcurve, which are likely related to the momentum dumps occurring during the observations. Using one of the aforementioned pipelines could lead to the inclusion of such candidates.\\
\indent In the future, the \textit{ESA PLATO} mission \citep{Rauer17} may result in high-precision light curves for a large sample of red giant stars from which Hot Jupiters can be discovered and their occurrence rate further constrained. 

\begin{acknowledgements}
    This work has made use of data from the European Space Agency (ESA) mission {\it Gaia} (\url{https://www.cosmos.esa.int/gaia}), processed by the {\it Gaia} Data Processing and Analysis Consortium (DPAC,
    \url{https://www.cosmos.esa.int/web/gaia/dpac/consortium}). Funding for the DPAChas been provided by national institutions, in particular the institutionsparticipating in the {\it Gaia} Multilateral Agreement. \\
    This paper includes data collected by the TESS mission. Funding for the TESS mission is provided by the NASA's Science Mission Directorate. \\
    This work has used the following additional software packages that have not been referred to in the main text: NumPy, Matpotlib, IPython and Pandas (\cite{NumPy};\cite{Matplotlib};\cite{IPython};\cite{Pandas}).
\end{acknowledgements}


\bibliographystyle{aa}
\bibliography{References.bib}

\clearpage
\newpage
\onecolumn
\begin{appendix}
\section{Gaussian processed lightcurves} \label{App:A}
\renewcommand{\thefigure}{A\arabic{figure}}

\begin{figure*}[h!]
    \centering
    \resizebox{0.7\textwidth}{!}{\includegraphics{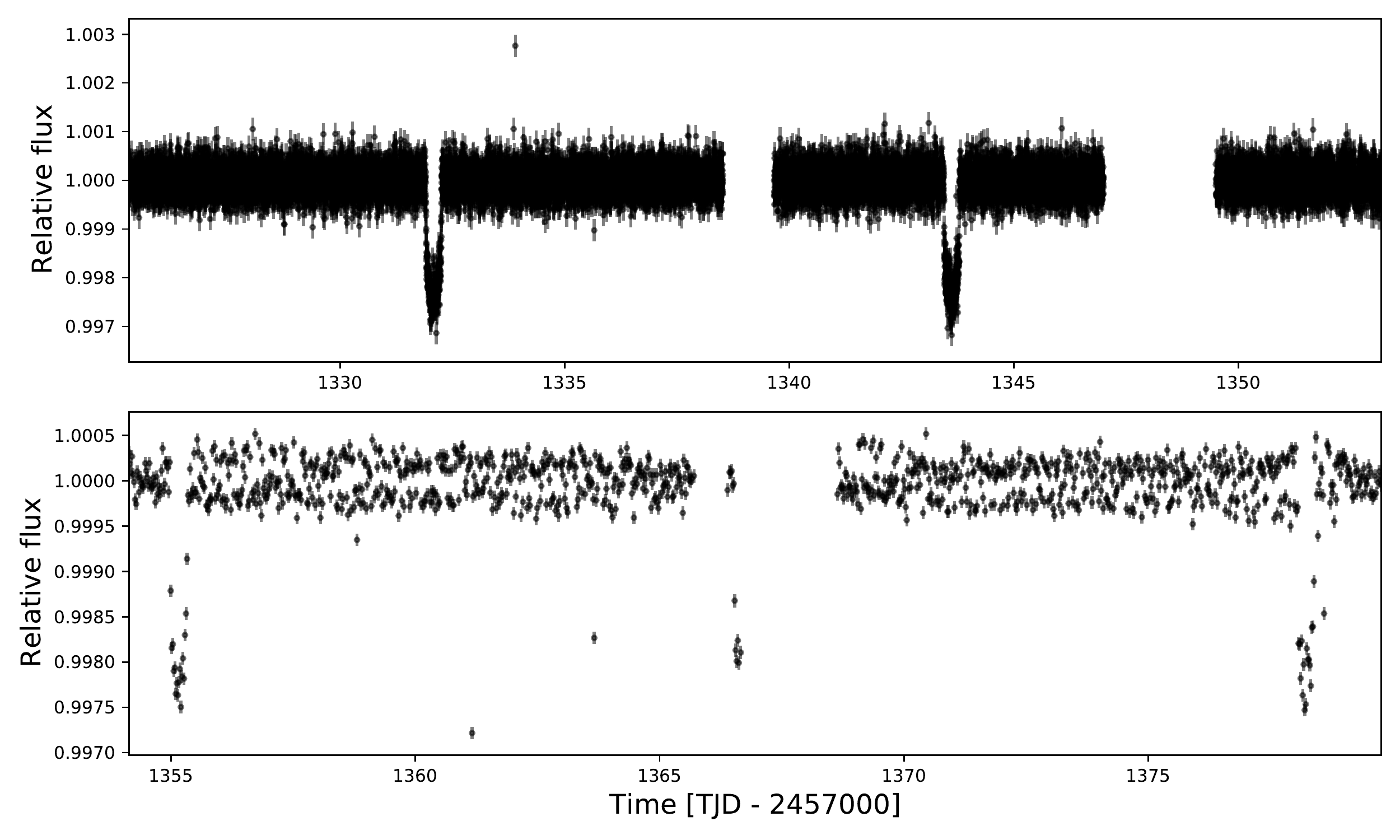}}
    \caption{The Gaussian processed lightcurve of HD 1397.}
    \label{fig:GP1397}
\end{figure*}
\begin{figure*}[ht!]
    \centering
    \resizebox{0.7\textwidth}{!}{\includegraphics{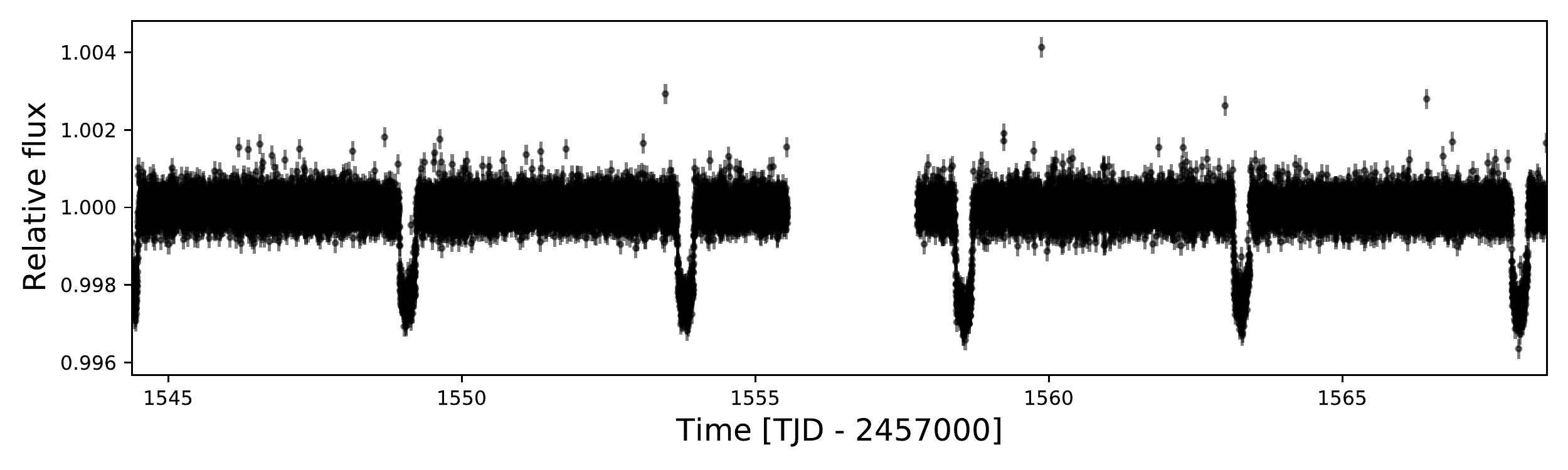}}
    \caption{As figure \ref{fig:GP1397}, but for the star KELT-11.}
    \label{fig:GP93396}
\end{figure*}
\begin{figure*}[ht!]
    \centering
    \resizebox{0.7\textwidth}{!}{\includegraphics{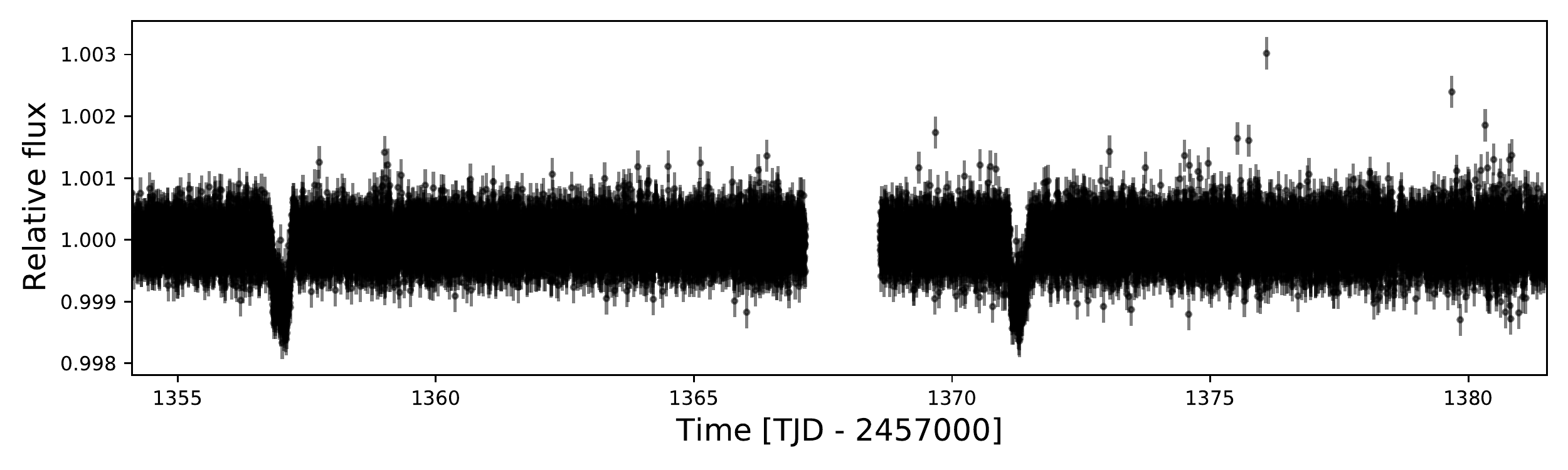}}
    \caption{As figure \ref{fig:GP1397}, but for the star HD 221416.}
    \label{fig:GP221416}
\end{figure*}
\begin{figure*}[ht!]
    \centering
    \resizebox{0.7\textwidth}{!}{\includegraphics{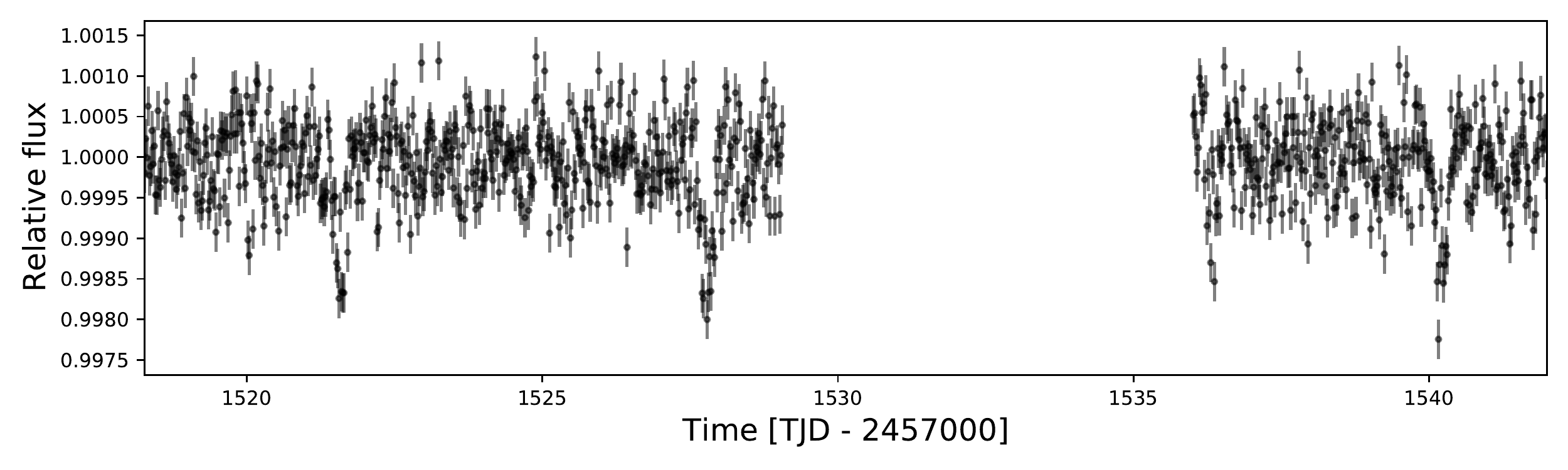}}
    \caption{As figure \ref{fig:GP1397}, but for the star TYC-5456-76-1.}
    \label{fig:GP5456}
\end{figure*}
\begin{figure*}[ht!]
    \centering
    \resizebox{0.7\textwidth}{!}{\includegraphics{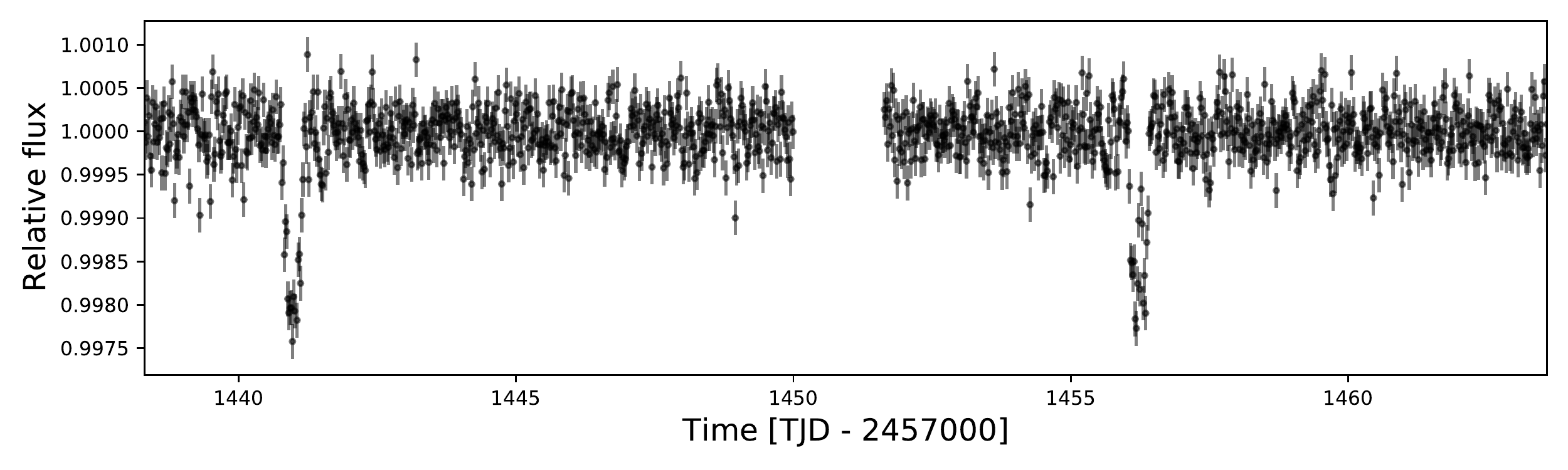}}
    \caption{As figure \ref{fig:GP1397}, but for the star CD-25 2180.}
    \label{fig:GP25}
\end{figure*}
\begin{figure*}[ht!]
    \centering
    \resizebox{0.7\textwidth}{!}{\includegraphics{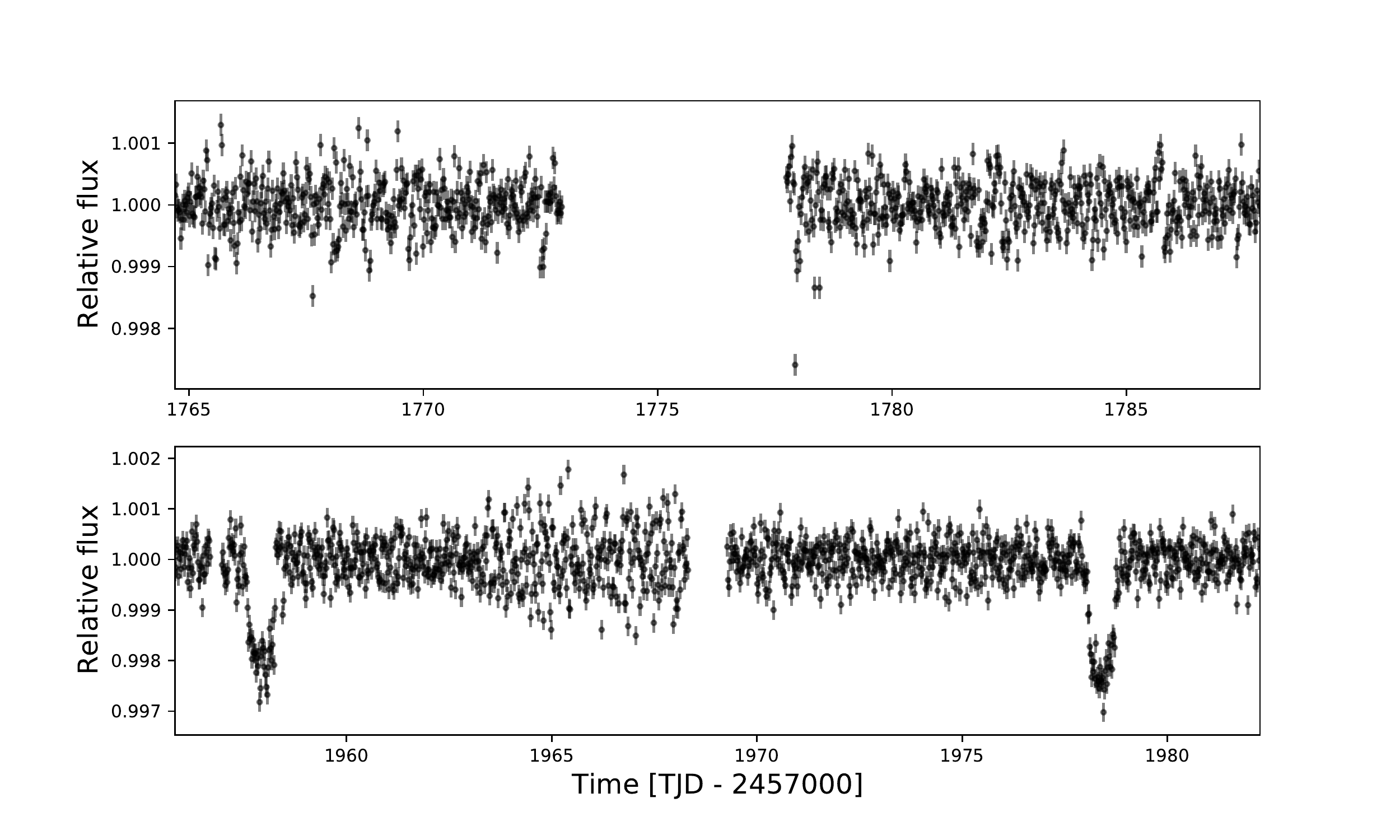}}
    \caption{As figure \ref{fig:GP1397}, but for the star TYC-4004-1387-1.}
    \label{fig:GP4004}
\end{figure*}

\clearpage
\section{Fitted lightcurves} \label{App:B}
\renewcommand{\thefigure}{B\arabic{figure}}
\begin{figure*}[ht!]
    \resizebox{\hsize}{!}{\includegraphics{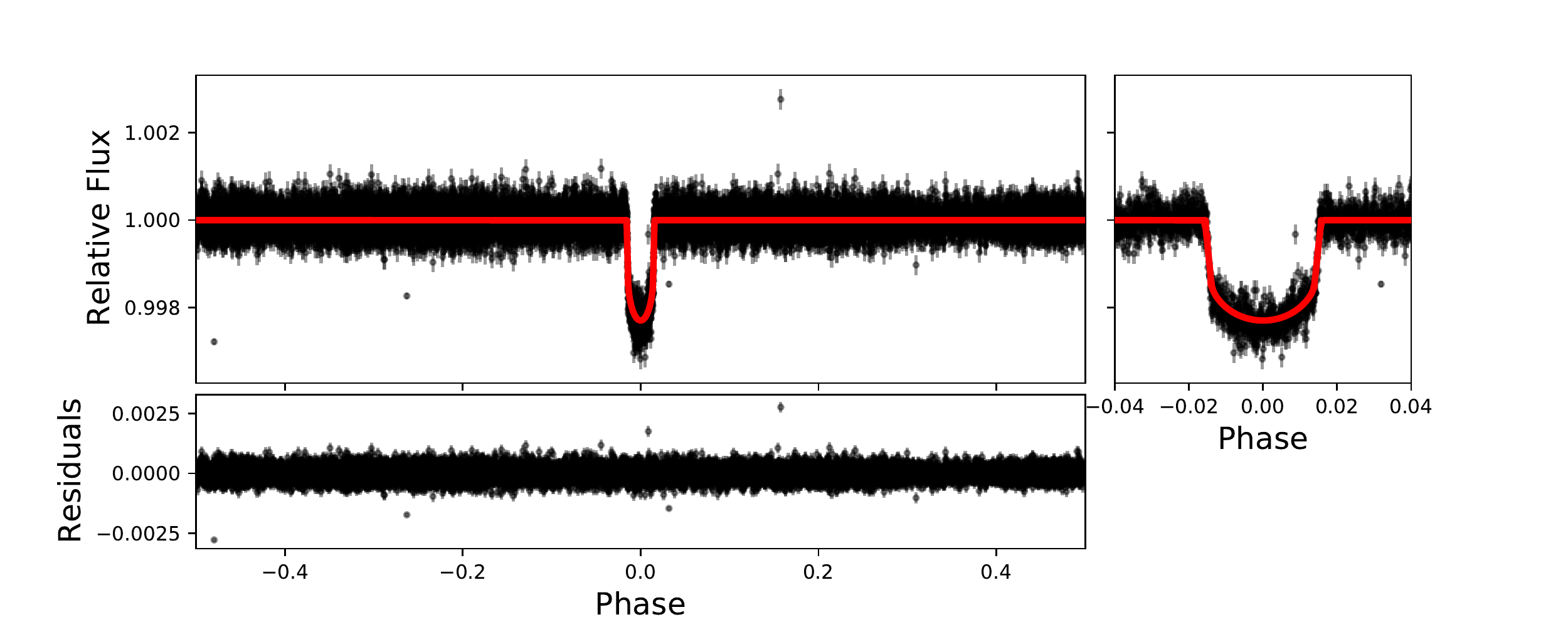}}
    \caption{The best fit to the lightcurve of HD 1397. The top left panel shows the phase-folded lightcurve with the best fit as the red line, while the bottom panel shows the residuals. The panel in the top right shows a zoomed-in version of the phase-folded transit.}
    \label{fig:Fit1397}
\end{figure*}
\begin{table*}[ht!]
\caption{The median values and $1\sigma$-confidence limits of the fitted parameters for the star HD 1397. }             
\label{table:Fit1397}      
\centering          
\begin{tabular}{c c c}     
\hline\hline       
Parameter & Median value & $1\sigma$-confidence intervals\\ 
\hline                    
    $P$ [days] & 11.53684 & $\pm$0.00022 \\
    $T_0$ [TJD] & 1332.08198 & $\pm$ 0.00045 \\
    $k$ & 0.04380 & (+0.00014, -0.00010) \\
    $a$ [$R$] & 10.72096 & (+0.04109, -0.12451) \\
    $i$ [degrees] & 89.54496 & (+0.31896, -0.50833) \\
    \hline
    $R_p$ [R$_\textnormal{Jup}$] & 0.98909 & (+0.04592, -0.05854) \\
\hline                  
\end{tabular}
\end{table*}

\begin{figure*}[ht!]
    \centering
    \resizebox{\hsize}{!}{\includegraphics{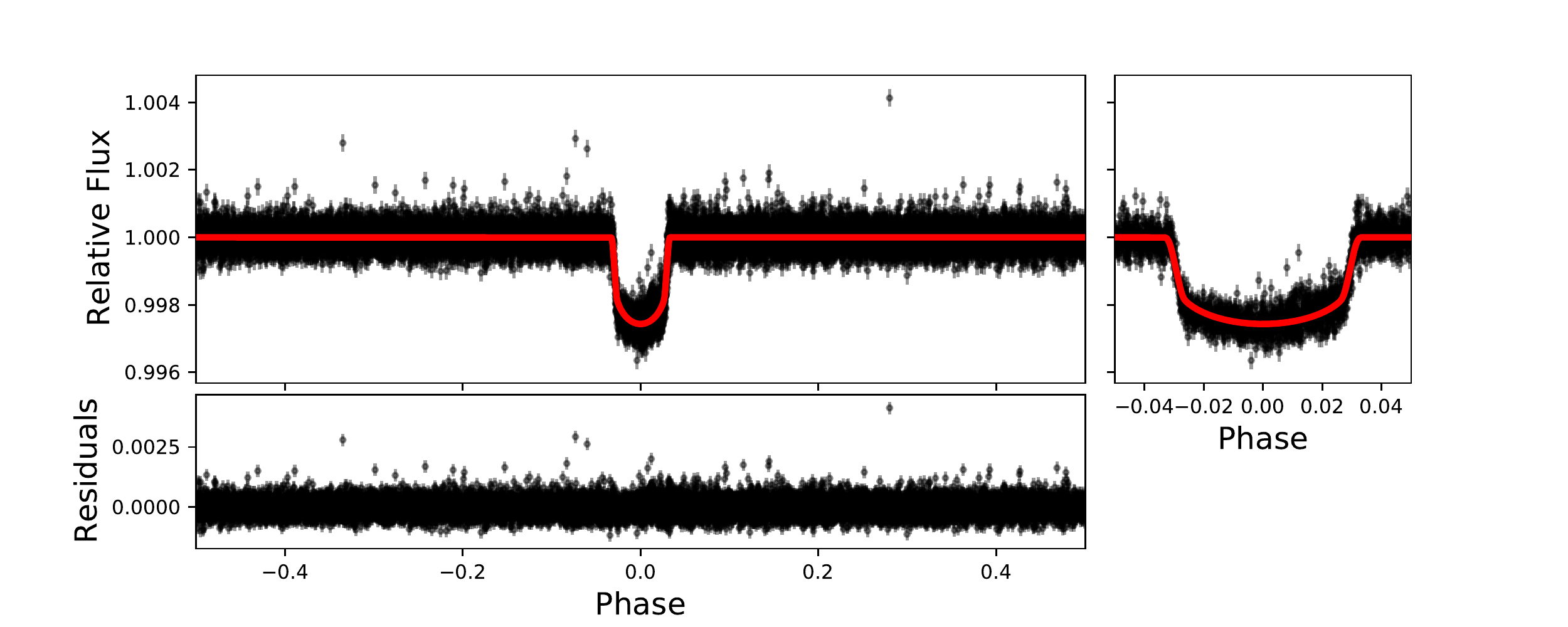}}
    \caption{As figure \ref{fig:Fit1397}, but for the star KELT-11.}
    \label{fig:Fit93396}
\end{figure*}
\begin{table*}[ht!]
\caption{As table \ref{fig:Fit1397}, but for the star KELT-11.}
\label{table:Fit93396}      
\centering          
\begin{tabular}{c c c}     
\hline\hline       
Parameter & Median value & $1\sigma$-confidence intervals\\ 
\hline                    
    $P$ [days] & 4.73554 & $\pm$ 0.00015 \\
    $T_0$ [TJD] & 1549.07760 & $\pm$ 0.00035 \\
    $k$ & 0.04664 & (+0.00047, -0.00028) \\
    $a$ [$R$] & 5.23613 & (+0.13231, -0.21427) \\
    $i$ [degrees] & 87.32559 & (+1.70869, -1.59199) \\
    \hline
    $R_p$ [R$_\textnormal{Jup}$] & 1.30042 & (+0.05684, -0.04012) \\
\hline                  
\end{tabular}
\end{table*}

\begin{figure*}[ht!]
    \resizebox{\hsize}{!}{\includegraphics{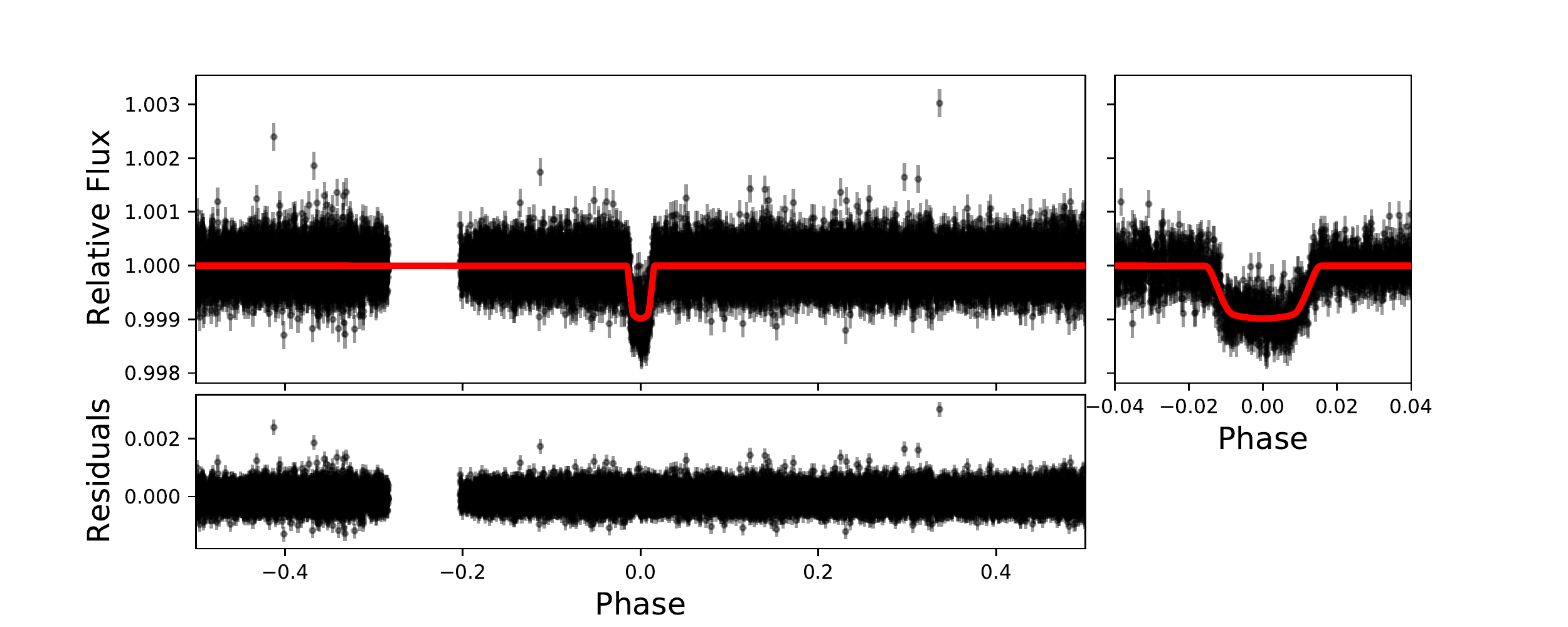}}
    \caption{As figure \ref{fig:Fit1397}, but for the star HD 221416.}
    \label{fig:Fit221416}
\end{figure*}
\begin{table*}[ht!]
\caption{As table \ref{fig:Fit1397}, but for the star HD 221416.}
\label{table:Fit221416}      
\centering          
\begin{tabular}{c c c}     
\hline\hline       
Parameter & Median value & $1\sigma$-confidence intervals\\ 
\hline                    
    $P$ [days] & 14.27848 & $\pm0.00354$ \\
    $T_0$ [TJD] & 1357.01034 & (+0.00234, -0.00242) \\
    $k$ & 0.03473 & (+0.00043, -0.00051) \\
    $a$ [$R$] & 4.80280 & (+0.37011, -0.25783) \\
    $i$ [degrees] & 78.81227 & (+0.94387, -0.73973) \\
    \hline
    $R_p$ [R$_\textnormal{Jup}$] & 1.00545 & (+0.03314, -0.04194) \\
\hline                  
\end{tabular}
\end{table*}

\begin{figure*}[ht!]
    \resizebox{\hsize}{!}{\includegraphics{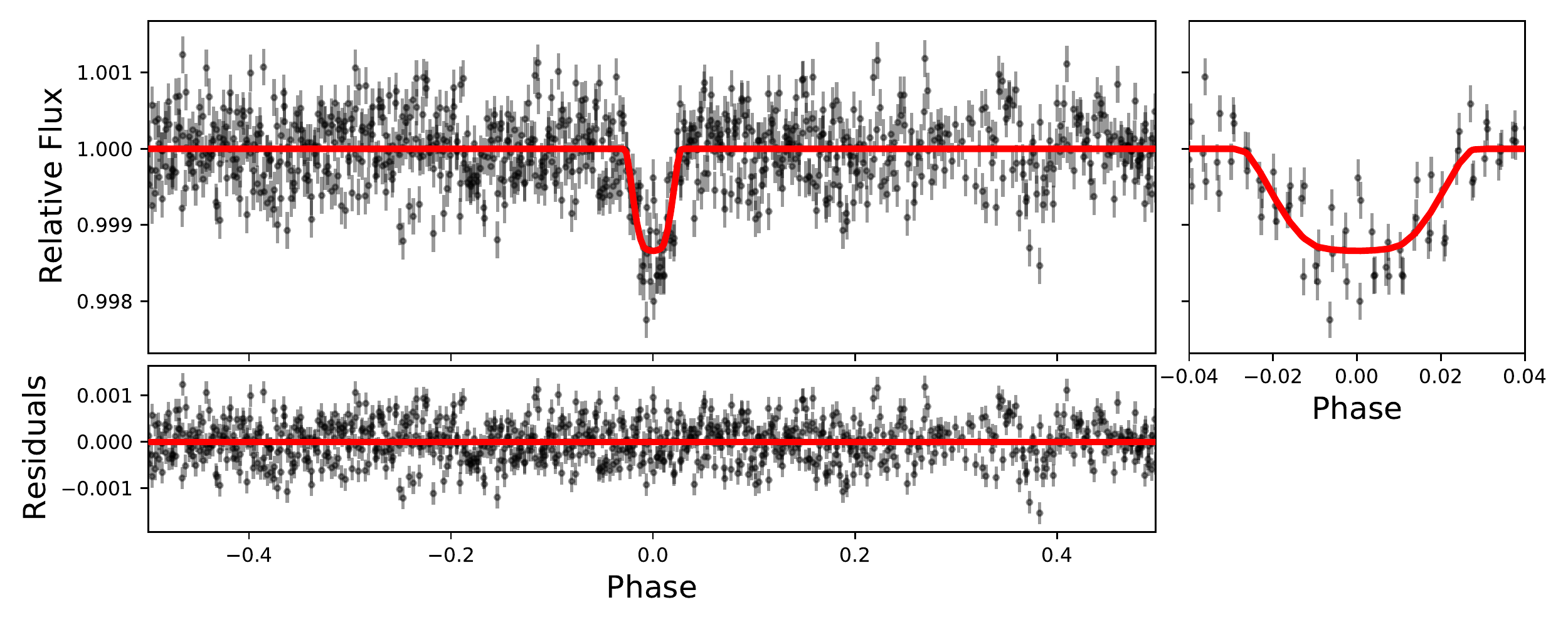}}
    \caption{As figure \ref{fig:Fit1397}, but for the star TYC-5456-1.}
    \label{fig:Fit5456}
\end{figure*}
\begin{table*}[ht!]
\caption{As table \ref{fig:Fit1397}, but for the star TYC-5456-76-1.}
\label{table:Fit5456}      
\centering          
\begin{tabular}{c c c}     
\hline\hline       
Parameter & Median value & $1\sigma$-confidence intervals\\ 
\hline                    
    $P$ [days] & 6.20965 & (+0.00270, -0.00284) \\
    $T_0$ [TJD] & 1521.57312 & (+0.00575, -0.00446) \\
    $k$ & 0.04129 & (+0.00454, -0.00858) \\
    $a$ [$R$] & 2.68558 & (+4.07250, -0.31119) \\
    $i$ [degree] & 69.32181 & (+19.06031, -3.35360) \\
    \hline
    $R_p$ [R$_\textnormal{Jup}$] & 1.58901 & (+0.17658, -0.33388) \\
\hline                  
\end{tabular}
\end{table*}

\begin{figure*}[ht!]
    \resizebox{\hsize}{!}{\includegraphics{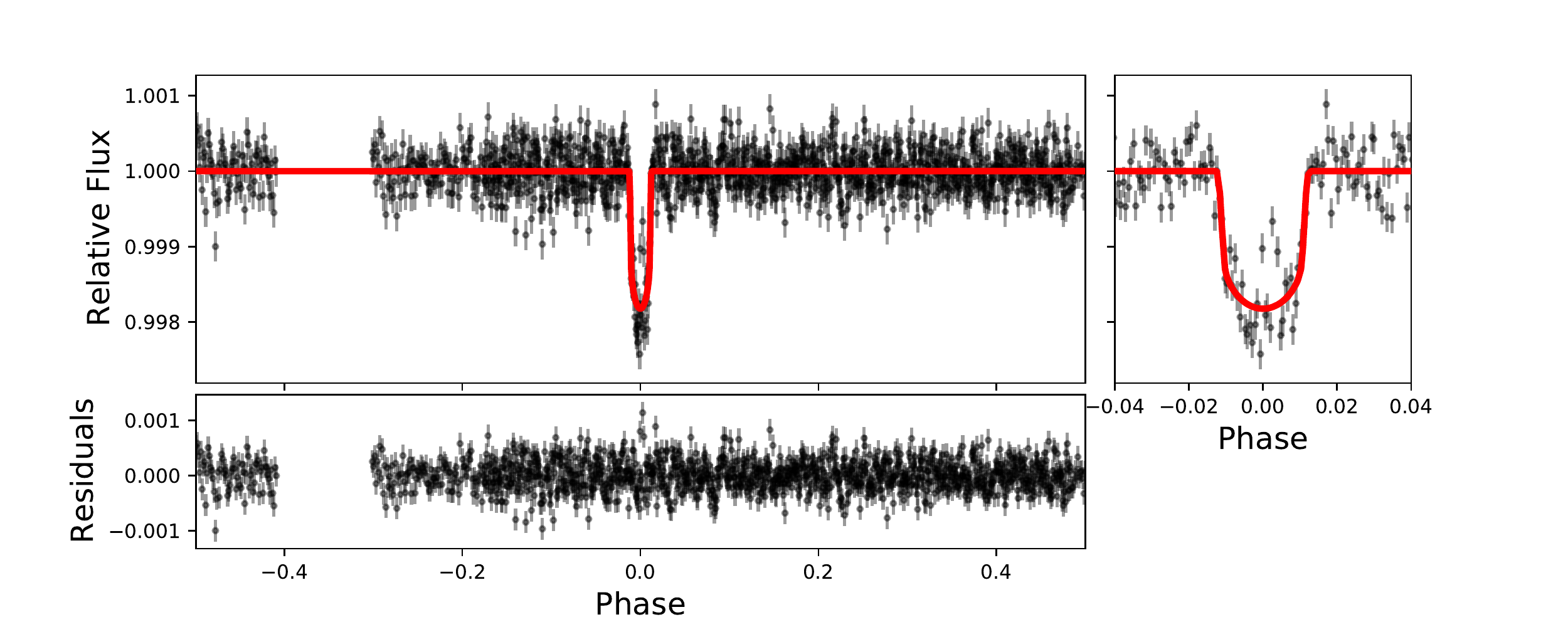}}
    \caption{As figure \ref{fig:Fit1397}, but for the star CD-25 2180.}
    \label{fig:Fit25}
\end{figure*}
\begin{table*}[ht!]
\caption{As table \ref{fig:Fit1397}, but for the star CD-25 2180.}
\label{table:Fit25}      
\centering          
\begin{tabular}{c c c}     
\hline\hline       
Parameter & Median value & $1\sigma$-confidence intervals\\ 
\hline                    
    $P$ [days] & 15.24236 & (+0.00431, -0.00420) \\
    $T_0$ [TJD] & 1440.98338 & (+0.00321, -0.00357) \\
    $k$ & 0.04072 & (+0.00304, -0.00151) \\
    $a$ [$R$] & 11.79079 & (+2.03693, -3.52985) \\
    $i$ [degree] & 87.30564 & (+1.96305, -2.98716) \\
    \hline
    $R_p$ [R$_\textnormal{Jup}$] & 1.56655 & (+0.11897, -0.06566) \\
\hline                  
\end{tabular}
\end{table*}

\begin{figure*}[ht!]
    \resizebox{\hsize}{!}{\includegraphics{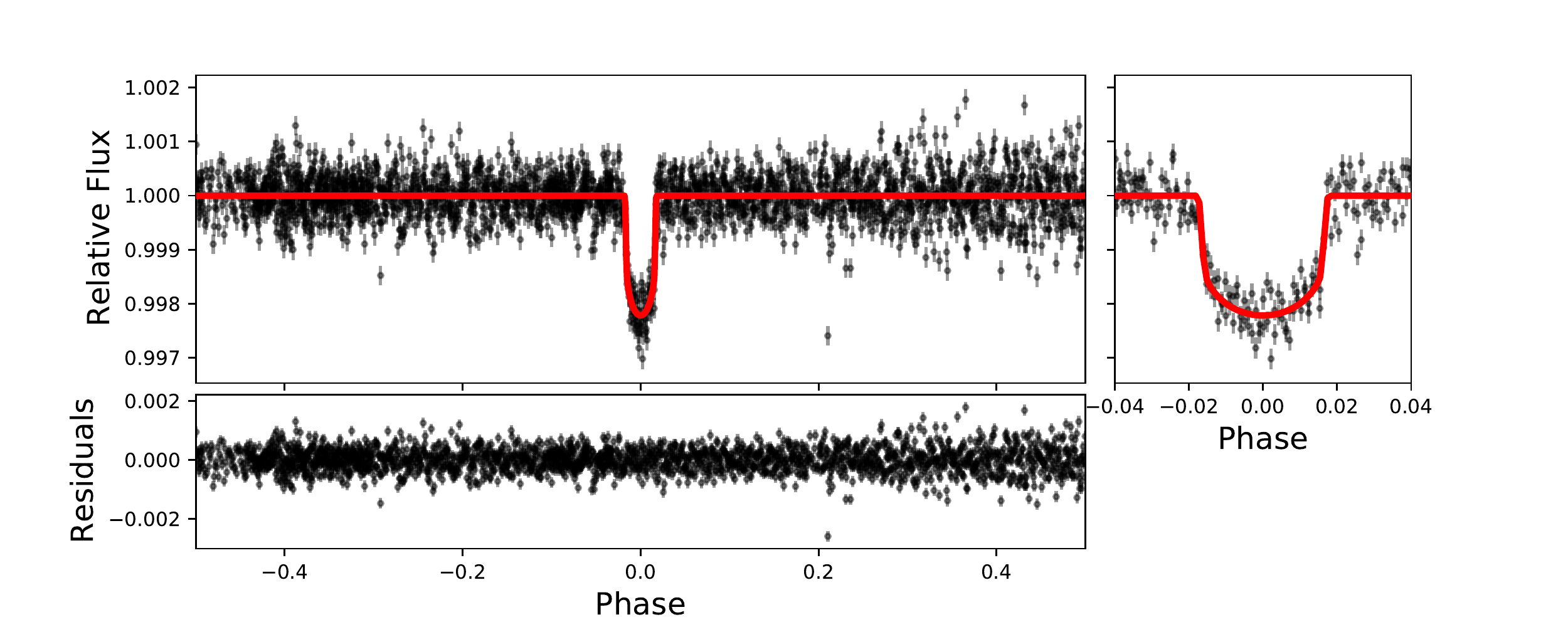}}
    \caption{As figure \ref{fig:Fit1397}, but for the star TYC-4004-1387-1.}
    \label{fig:Fit4004}
\end{figure*}
\begin{table*}[ht!]
\caption{As table \ref{fig:Fit1397}, but for the star TYC-4004-1387-1.}
\label{table:Fit4004}      
\centering          
\begin{tabular}{c c c}     
\hline\hline       
Parameter & Median value & $1\sigma$-confidence intervals\\ 
\hline                    
    $P$ [days] & 20.47783 & (+0.00364, -0.00359) \\
    $T_0$ [TJD] & 1957.92256 & (+0.00253, -0.00258) \\
    $k$ & 0.04328 & (+0.00068, -0.00039) \\
    $a$ [$R$] & 9.36247 & (+0.22352, -0.63576) \\
    $i$ [degree] & 88.60308 & (+0.97259, -1.38418) \\
    \hline
    $R_p$ [R$_\textnormal{Jup}$] & 1.59120 & (+0.06655, -0.06935) \\
\hline                  
\end{tabular}
\end{table*}
\end{appendix}

\end{document}